\documentclass[preprint2]{aastex}

\shorttitle{On the Gas Temperature of Molecular Cloud Cores}
\shortauthors{Juvela et al.}

\begin{document}

\title{On The Gas Temperature of Molecular Cloud Cores}

\author{M. Juvela and N. Ysard}
\affil{Department of Physics, University of Helsinki, FI-00014 Finland}

\begin{abstract}

We investigate the uncertainties affecting the temperature profiles of
dense cores of interstellar clouds. In regions shielded from external
ultraviolet radiation, the problem is reduced to the balance between
cosmic ray heating, line cooling, and the coupling between gas and
dust. We show that variations in the gas phase abundances, the grain
size distribution, and the velocity field can each change the
predicted core temperatures by one or two degrees. We emphasize the
role of non-local radiative transfer effects that often are not taken
into account, for example, when modelling the core chemistry. These
include the radiative coupling between regions of different
temperature and the enhanced line cooling near the cloud surface. The
uncertainty of the temperature profiles does not necessarily translate
to a significant error in the column density derived from
observations. However, depletion processes are very temperature
sensitive and a two degree difference can mean that a given molecule
no longer traces the physical conditions in the core centre.

\end{abstract}


\section{Introduction}

The gas kinetic temperature $T_{\rm gas}$ is a central parameter for
the stability of the clouds, their evolution, and their chemistry
\citep[e.g.][]{Galli02, Keto05}. The properties of pre-stellar cores
are of particular interest because they have a direct impact on the
ensuing star formation. The understanding of the temperature
variations within the cores is thus crucial for the interpretation of
observations and for the development of the theory of star formation
\citep{Bergin2007}.

The early models of the cloud cores were spherically symmetric and
isothermal, in the case of both stable and collapsing cores 
\citep[e.g.][]{Bonnor56,Shu77}. The spherical symmetry will eventually
break down because of rotational flattening or core fragmentation but,
in early evolutionary stages, it is often a reasonable first
approximation \citep[e.g.][]{Evans05,Alves01}. The observations of the
detailed temperatures structure can be more difficult because the
measured intensities are non-linear averages of the emission
originating in different regions along the line-of-sight. The dust
colour temperature is known to be a biased estimate of the true grain
temperature $T_{\rm dust}$ \citep[][]{Shetty09,Malinen11}. The
situation is potentially even worse for $T_{\rm gas}$ where, because
of abundance variations, the observations may represent conditions in
a small fraction of the source. This is particularly true for cold
pre-stellar cores where many molecules are severely depleted
\citep[e.g.][]{Bergin97, Belloche04, Whittet10, Ford11, Parise11}.
There are measurements suggesting that, in the centre of a starless
core, the kinetic temperature can decrease down to $\sim$6\,K
\citep{Crapsi07,Harju08}. However, our knowledge of the radial
temperature variations depends heavily on models that also form the
basis for the interpretation of molecular line data.

The principles of the thermal balance of dense clouds are well
understood, the main factors being the heating by cosmic rays, the
cooling by line emission, and the energy exchange between gas and dust
\citep{Goldsmith78, Goldsmith01}. Despite the apparent simplicity of
the problem, reliable predictions of $T_{\rm gas}$ are not easy to
obtain because of the uncertainty of the fractional abundances and
dust grain sizes, and the potentially complex radiative transfer
effects. The problem of line cooling has been examined separately,
with the Large Velocity Gradient (LVG) approximation \citep{Neufeld95}
and with Monte Carlo methods \citep{Juvela01}, the latter also
enabling the study of the role of an inhomogeneous medium.

In this paper, we investigate the uncertainties of the modelled
$T_{\rm gas}$ profiles of dense cloud cores. The cores are known or
are suspected to have significant radial variations in the gas phase
abundances, grain size distributions, and the velocity field. We wish
estimate the maximum effects on $T_{\rm gas}$ that could arise from
these systematic changes. This is important for the evaluation of the
uncertainties of chemical model and, more directly, the interpretation
of any line observations of dense cores.

We describe our models in Sect.~\ref{sect:model}. The results are
presented in Sect.~\ref{sect:results}, both for homogeneous models
(Sect.~\ref{sect:goldsmith}) and for a Bonnor-Ebert type cloud
(Sect.~\ref{sect:BE}), and our conclusions are presented in
Sect.~\ref{sect:conclusions}.

\section{The Modelling} \label{sect:model}

We examine the thermal balance of spherical clouds without small scale
inhomogeneity. All models are divided to one hundred concentric
shells, the innermost shell being 6\% and the outermost shell 0.6\% of
the outer radius. The same discretization is used for in continuum and
line radiative transfer calculations and for the determination of the
temperature profiles.  
To separate the radiative transfer effects from those
of the density, we start by examining homogeneous clouds. The dust
temperatures are determined with Monte Carlo radiative transfer
calculations \citep{Juvela05}, using the dust model of
\citet{Draine03} and the interstellar radiation field (ISRF) given by
\citet{Black94}. 
The rate for the energy exchange between gas and dust
is calculated from
\begin{eqnarray}
\Lambda_{g,d} = 2 \times 10^{-33} [ n(H_2)/cm^{3} ]^2
   (T_{\rm gas}-T_{\rm dust}) 
 \nonumber \\
   (T_{gas}/10.0\,K)^{0.5}  
   \, erg \,cm^{-3} \,s^{-1}.
   \label{eq:gd}
\end{eqnarray}
and the rate for cosmic ray heating from
\begin{equation}
\Gamma_{gas, cr} = 10^{-27} [ n(H_2)/cm^{3} ] \, erg\, cm^{-3} \,
s^{-1}
\end{equation}
\citep{Goldsmith01}. Following \cite{Goldsmith01}, we calculate the
line cooling $\Lambda_{\rm gas}$ by $^{12}$CO, $^{13}$CO, C$^{18}$O,
C, CS, and o-H$_2$O, multiplying the last two rates by 10 and two,
respectively, to take into account the cooling by other species. The
radiative transfer problem is solved with Monte Carlo methods
\citep{Juvela01}, with abundances given in Table 1 in
\citet{Goldsmith01}. As part of the Monte Carlo simulation, the
program saves the net radiative cooling rates for each model cell.

As a more realistic cloud model we examine a one solar mass,
marginally stable Bonnor-Ebert sphere with a value of the stability
parameter $\xi=6.5$ \citep{Bonnor56}. The radial density profile is
calculated assuming a constant temperature of 10\,K. The variations of
kinetic temperature modify the density profile only slightly
\citep[][]{Evans01} and this will not affect the conclusions drawn
from the models.
An external UV field can affect the cloud temperatures but, in
principle, only in a thin surface layer. We consider this heating
through the photoelectric effect in connection with some Bonnor-Ebert
models. To estimate the photoelectric heating rate $\Gamma_{\rm PE}$,
we first calculate the energy that dust absorbs in the energy range
6--13.6\,eV. The values are obtained from the continuum radiative
transfer calculations. Following \cite{Juvela03}, $\Gamma_{\rm PE}$ is
obtained by multiplying these energies with an constant efficiency of
$\epsilon=0.029$. 
The dust temperatures are calculated assuming the clouds are heated by
the normal ISRF, even when UV heating of gas is ignored. 
Photoelectric heating is almost completely eliminated if the cloud is
surrounded by a dust layer of $A_{\rm V}\sim 2^{\rm m}$ (see
Sect.~\ref{sect:BE_T}). This would reduce the central dust
temperature only by $\sim$0.5 degrees, less than the uncertainty
associated with the selection of a dust model.

Below we modify several model parameters to determine how their
variations are reflected on the gas temperature. These include (1) the
thermal coupling between gas and dust, $\Lambda_{g,d}$, that depends
on the grain size distribution, (2) the abundances of the cooling
species that depend on the degree of depletion, (3) the cosmic ray
heating $\Gamma_{gas, cr}$ that depends on the rate of cosmic rays,
and finally (4) the large scale infall motion and (5) the small scale
velocity field that both affect the radiative cooling $\Lambda_{\rm
g}$. The degree of modification for each parameter is indicated in
Table~\ref{table:parameters} and discussed further in the following
section.

\section{Results} \label{sect:results}

\subsection{Homogeneous models} \label{sect:goldsmith}

\begin{figure}
\plotone{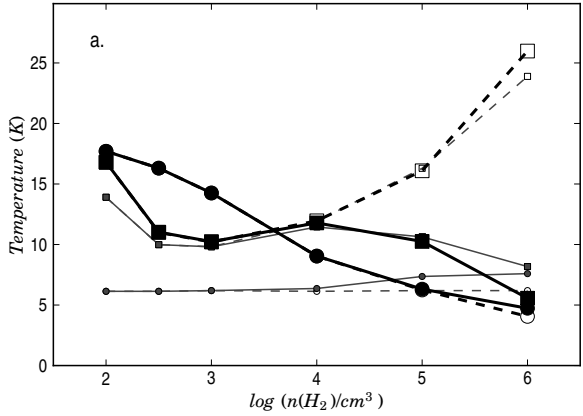}
\caption{
A comparison of the $T_{\rm gas}$ (squares) and $T_{\rm dust}$
(circles) in our calculations (large symbols) and in
\citet{Goldsmith01} (small symbols). The open symbols denote the
results without the gas--dust coupling.
\label{fig:goldsmith}}
\end{figure}

As a first test, before modifying any parameters listed in
Table~\ref{table:parameters}, we compared our calculations with those of
\citet{Goldsmith01} who estimated the line cooling with LVG modelling
assuming a velocity gradient of 1\,km\,s$^{-1}$\,pc$^{-1}$. The cloud
is taken to be well shielded from the external UV field so that the
heating through the photoelectric effect can be neglected.
Corresponding to the idea of the LVG models, we calculated $T_{\rm
gas}$ at the centre of homogeneous, microturbulent spheres with a
radius of $R=1$\,pc and the line Doppler width equal to
1\,km\,s$^{-1}$. The correspondence was found to be good, mostly
within $\sim$10\% in $T_{\rm gas}$ (see Fig.~\ref{fig:goldsmith}a), in
spite several basic differences in the respective models. Firstly,
\cite{Goldsmith01} calculated the dust temperature for a constant
shielding that, in the absence of gas-dust coupling resulted in a
temperature of $T_{\rm dust}\sim$6\,K. In our calculations the dust
temperature was solved self-consistently. The model was assumed to be
illuminated by the full ISRF, and the radiation field inside the cloud
was solved with radiative transfer calculations \citep{Juvela05}. The
difference in dust temperature has no effect on $T_{\rm gas}$ at low
$n({\rm H_2})$. At $n({\rm H_2})=10^5$\,cm$^{-3}$ $T_{\rm dust}$ is
close to the value used by \citet{Goldsmith01} and the $T_{\rm gas}$
values are in very good agreement. Secondly, in our model the
excitation temperature $T_{\rm ex}$ varies as a function of radius so
that the LVG assumption of a uniform medium is not valid. This affects
the radiative connection between different parts of the model, is
reflected in the excitation, and can thus affect the cooling rates.
Thirdly, the photon escape probabilities in the LVG model and our
Monte Carlo model are not identical even when the total optical depths
are equal. In our calculations, even at high optical depths, some
photons can always escape in the line wings. These effects do not
appear important for the estimated central temperature of the cores.
There are also some differences. For example, in our calculations the
gas-dust coupling has a smaller effect on $T_{\rm dust}$ at high
densities, apparently because our dust cooling rate is higher than
that given by \citet{Goldsmith01} Eq. 13.

We used the homogeneous models to investigate the effect of the grain
size distribution. The gas--dust coupling becomes significant around
$n({\rm H}_2)\sim 10^5$\,cm$^{-3}$ depending, however, on the total
grain area. Equation.~\ref{eq:gd} is valid for a size distribution
$dn/da~\sim a^{-3.5}$ with $a$ in the range 0.01--1.0\,$\mu$m. If the
lower limit is reduced to 10\AA\, \citep[e.g.][]{Li01}, without
modifying the gas-to-dust ratio, $\Lambda_{g,d}$ increases by a factor
of $\sim$3. On the other hand, at the centre of dense cores the size
of the large grains increases through grain coagulation while small
grains may disappear entirely \citep[e.g.][]{Stepnik03, Ormel09,
Steinacker10}. If the lower limit of grain sizes increases to 500\AA,
the rate $\Lambda_{g,d}$ is reduced by 60\%. If the upper limit is
further increased to 2\,$\mu$m, the effect is a factor of three. The
value of $\Lambda_{g,d}$ is similarly increased (decreased) by a
factor of three if the powerlaw exponent of the size distribution is
decreased (increased) by $\sim0.75$, without modifying the size
limits.

\begin{figure}
\plotone{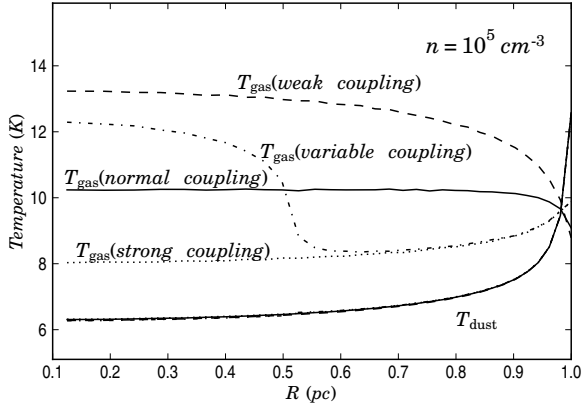}
\caption{
The radial profiles of $T_{dust}$ (lower solid line) and
$T_{\rm gas}$ (upper solid line) for a homogeneous cloud with a density
of $n=10^5$\,cm$^{-3}$. The other lines correspond to three times
stronger or three times weaker coupling or a combination of the two
where the coupling is weaker in the centre.
\label{fig:radial}}
\end{figure}

Figure~\ref{fig:radial} illustrates the consequences for the gas
temperature. The solid lines show the radial profiles of $T_{\rm gas}$
and $T_{\rm dust}$ for the model $n=10^5$\,cm$^{-3}$ of
Fig.~\ref{fig:goldsmith}. The increased photon escape probability
always decreases $T_{\rm gas}$ at the cloud surface in spite of the
increasing $T_{\rm dust}$ (Eq.~\ref{eq:gd}). The other curves
correspond to a three times stronger and a three times weaker gas-dust
coupling. The dash-dotted line is schematically the expected behaviour
where, for given density, the coupling becomes weaker in the central
part because of the grain growth. This is not to be taken so much as a
model of an actual core as an illustration of the uncertainty of
$T_{\rm gas}$.

\subsection{Bonnor-Ebert spheres}  \label{sect:BE}

As more realistic models of dense cores, we examine critically stable
0.5 and 1.0 solar mass Bonnor-Ebert spheres. We will first investigate
different factors that could affect their radial temperature profiles
and then have a look at the predicted line profiles.

\subsubsection{Temperature profiles} \label{sect:BE_T}

We investigate first the one solar mass Bonnor-Ebert sphere. The basic
model has a constant turbulent line width of $\sigma_{\rm
V}=1$\,km\,s$^{-1}$ and no large scale gas motions.  The central
density rises above $10^5$\,cm$^{-3}$ but, because of the smaller
cloud size, the column densities are lower than in the
Sect.~\ref{sect:goldsmith} models of similar density. The photon
escape probability increases outwards and the model predicts a
significant decrease of $T_{\rm gas}$ towards the cloud surface. This
is in contrast with the \citet{Galli02} calculations that employed a
parameterization of LVG results to estimate $\Lambda_{gas}$ (see their
Fig. 3; note also the difference in $T_{\rm dust}$ due to a different
dust model). The actual surface temperature will be sensitive to the
abundance profiles and will depend on the amount of UV and cooling
line radiation entering the cloud from the outside, both effects
ignored in the present model.

Figure~\ref{fig:BE_1.0} shows the quantitative effects resulting from
possible variations of the abundances, the velocity field, and
$\Lambda_{g,d}$. The question of depletion was already examined by
\citet{Goldsmith01}. To illustrate the effect in the context of our
model, we decrease the abundance of all cooling species by a factor of
ten in the cloud centre, within a radius of 0.025\,pc
(Fig.~\ref{fig:BE_1.0}a). Because many lines are already optically
thick, $\Lambda_{\rm g}$ is not expected to decrease linearly with the
abundances. In the model, $T_{\rm gas}$ increases by $\sim$2 degrees
in the inner part, with only a small effect reflected in the outer
cloud.

In quiescent cores the line widths are sometimes observed to be close
to that determined by thermal broadening \citep[e.g.][]{Harju08}. When
the turbulent line width is reduced to $\sigma_{\rm
V}=0.1$\,km\,s$^{-1}$ within innermost 0.025\,pc, the central
temperature again increases by about two degrees 
(Fig.~\ref{fig:BE_1.0}b). The observed linewidths will remain much
broader because of the thermal and opacity broadening and because of
the emission from the outer cloud layers where the turbulent line width
is still $\sigma_{\rm V}=1.0$\,km\,s$^{-1}$.

Although the Bonnor-Ebert model is static, we can introduce a large
scale velocity field to check its importance on the escape of line
emission. We add an infall velocity that is zero at the cloud surface
and increases linearly to 1\,km\,s$^{-1}$ in the centre. This is a very
simplistic model of the velocity field but the magnitude of the
velocity gradient is realistic \citep[e.g.][]{Zhou93} and the model
should capture the main effect on the radiative transfer. In the very
central part of the model, the gas temperature by reduced little less
than one degree (Fig.~\ref{fig:BE_1.0}c). Both the large scale and
small scale velocity fields affect $\Lambda_{\rm g}$ through the line
optical depths. 

The density of the one solar mass model ranges from 1.4$\times 10^4$
to 2$\times 10^5$\,cm$^{-3}$, a region of densities where the coupling
between gas and dust becomes important. As discussed in
Sect.~\ref{sect:goldsmith}, changes in the grain size distribution are
reflected on the efficiency of $\Lambda_{g,d}$ so that it could be
decreased by up to a factor of three in the cloud centre and enhanced
by up to a factor of three at the surface. It may be very improbable
that a single source would exhibit the full range of variation.
However, Fig.~\ref{fig:BE_1.0}d shows this case where the efficiency
of $\Lambda_{g,d}$ jumps from one extreme to the other again at
0.025\,pc radius. Because the dust temperature is {\em above} the gas
temperature, the weaker coupling reduces $T_{\rm gas}$ in the centre
of the cloud. The effect is again of the order of one degree.

Figure~\ref{fig:BE_0.5} shows the corresponding results for a half
solar mass cloud where the central density is four times the value of
the previous model. The main differences result from the enhanced
gas-dust coupling that, together with lower dust temperature, reduces
the central gas temperature by $\sim 1$\,K in the cases of low
abundances and low velocity dispersion. On the other hand, the outer
cloud is warmer by a similar amount, both because of the stronger
$\Lambda_{g,d}$ but also because of the general density dependence
already seen in Fig.~\ref{fig:goldsmith}.

An external UV field can directly impact the cloud temperature at
least at its surface. We examined another set of one solar mass
Bonnor-Ebert spheres where the photoelectric heating was included and
calculated as described in Sect.~\ref{sect:model}. The ISRF impinging
on the cloud is attenuated by $A_{\rm V}$=0, 1, or 2$^{\rm m}$,
corresponding to a shielding dust layer that is thought to exist
around the actual model cloud. This applies to the calculation of both
$T_{\rm dust}$ and $\Gamma_{\rm PE}$. The resulting temperature
profiles inside the model are shown in Fig.~\ref{fig:PEH}. The effect
of photoelectric heating becomes negligible once the cloud is shielded
by $A_{\rm V}\sim 2^{\rm m}$. However, in an unshielded cloud the UV
field has a small effect, $\sim$0.4\,K, even in the cloud centre. This
is caused not by a direct photoelectric heating but by the change of
the excitation in the outer cloud layers (cf.
Sect.~\ref{sect:nonlocal}, Fig.~\ref{fig:goldsmith}b). Again, the
actual effect will depend critically on the molecular abundances in
the region heated by $\Gamma_{\rm PE}$.

As a final source of uncertainty we consider the rate of cosmic rays.
The heating term $\Gamma_{gas, cr}$ is based on the assumption of a
rate $\zeta=3\times 10^{-17}$\,s$^{-1}$ but, in diffuse clouds, there
are reports of rates that are up to two orders of magnitude higher 
\citep[e.g.][]{McCall03, Liszt03, Shaw06}. To reflect this uncertainty
we calculated temperature profiles with $\zeta$ scaled by factors 1,
2, 5, and 10. Figure~\ref{fig:CR} shows the resulting temperature
profiles, again for the one solar mass model. With the highest rate,
$\zeta=3\times 10^{-16}$\,s$^{-1}$, the central temperature has risen
from the original $\sim$7.5\,K to $\sim$17\,K. For comparison, we show
temperatures calculated with the $\Lambda_{gas}$ parameterization given
by \citet{Goldsmith01}. The calculation is done shell by shell using
the local density and the local dust temperature ($\Gamma_{cr}$ and
$\Lambda_{gas, dust}$ are as in our calculations). These
$\Lambda_{gas}$ rates correspond to a model that has much higher
column density (per velocity interval and for given density) than the
Bonnor-Ebert spheres. Therefore, also the derived $T_{\rm gas}$ values
are higher. Furthermore, the parameterization does not catch the
increased photon escape probability at the cloud surface that, in our
Monte Carlo calculations, results in the decrease of temperature in
the outer part.

\subsubsection{Spectral lines}

Figures~\ref{fig:spectra_1.0} and \ref{fig:spectra_0.5} show $^{13}$CO
and C$^{18}$O line profiles that were calculated for the models of
Figs.~\ref{fig:BE_1.0} and \ref{fig:BE_0.5}. The $J=$1--0, $J=$2--1,
and $J=$3--2 spectra were calculated as observed towards the centre of
the cloud with a beam with the FWHM equal to the half of the cloud
radius. At 100\,pc distance, this corresponds to $\sim$52\arcsec and
$\sim$26\arcsec for the 1.0\,$M_{\sun}$ and the 0.5\,$M_{\sun}$
models, respectively.

In the one solar mass model, the $^{13}$CO(1--0) beam averaged optical
depth is only $\tau \sim$2.5 through the cloud. In the 0.5\,$M_{\sun}$
the corresponding optical depth is $\sim$7.5 meaning that there the
$^{13}$CO transitions and C$^{18}$O lines originate partially in
different regions with different kinetic temperatures. One example of
this are the line profiles of the 0.5\,$M_{\sun}$ cloud with the
infall velocity. The $^{13}$CO(2-1) line shows the expected infall
profile while in the optically thinner $J=$1-0 line ($\tau$=3.9 vs
$\tau$=6.5 for the second transition) the effect is weaker.  The beam
averaged C$^{18}$O(1--0) lines remain symmetric. When observed with a
pencil beam, the C$^{18}$O $J$=1--0 and $J$=2--1 lines would show
slight asymmetry but with line profiles with stronger emission on the
red shifted side. This is caused by the $T_{\rm gas}$ which decreases
towards the cloud centre (see Fig.~\ref{fig:BE_0.5}c).

We carried out LTE analysis of the $^{13}$CO and C$^{18}$O lines to
check how accurate those column density estimates would be. We used
the method described by \citet{Myers83}. The ratio of the $^{13}$CO
and C$^{18}$O lines is used to calculate the optical depth
$\tau_{18}$, and the excitation temperature of C$^{18}$O,
$T^{18}_{ex}$, is solved from the radiative transfer equation 
\citep[see][Eqs. 3-4]{Myers83}. Assuming Gaussian line shapes, the
column density of C$^{18}$O in the $J=1$ state is 
\begin{equation}
N_{J=1} = 3.86 \times 10^{14} \tau_{18} J(T^{18}_{\rm ex}) \Delta v_{18}
\; {\rm cm}^{-2}
\end{equation}
where $J(T) = T_0 / [exp(T_0/T) - 1 ]$ and $T_0 = 5.27$ K.  The total
C$^{18}$O column density is obtained by summing all levels, assuming
they are populated according to $T^{18}_{ex}$.
Table~\ref{table:LTE} summarizes the results when the line
parameters are taken from Gaussian fits to the modelled spectral
profiles. The errors of these estimates are less than 5\% which shows
that the temperature gradients have little impact on the column
density estimates.

\section{The importance of non-local radiative couplings}
\label{sect:nonlocal}

The LVG method is based on the assumption that the excitation is
constant within the radiatively coupled volume. This is not a good
approximation in dense cores where the gas velocities are small
and the radial gradients of $T_{\rm ex}$ are large. To illustrate the
potential problem further in a schematic way, we took from
Fig.~\ref{fig:goldsmith} the $n=10^5$\,cm$^{-3}$ model and forced the
inner part, $r<0.5$\,pc, to local thermodynamic equilibrium (LTE) at
5\,K or 20\,K. Fig.~\ref{fig:LTE} shows the resulting $T_{\rm
gas}$ values in the outer part of the cloud, $r>0.5$\,pc. The assumed
excitation of the inner cloud has a strong impact on the remaining
cloud volume. The same effect was seen in Sect.~\ref{sect:BE}, where
the increased surface temperature caused by photoelectric heating was
reflected all the way to the centre of the optically thick cloud.

Our final example of a system with non-trivial radiative couplings
consists of two Bonnor-Ebert spheres with properties identical to
those in Sect.~\ref{sect:BE}.  The spheres are touching each other and
the mutual shielding and exchange of radiative energy modifies the
temperature distributions. Figure~\ref{fig:two_spheres} shows the
$T_{\rm gas}$ estimated without and with photoelectric heating. The
calculations were carried out with the same continuum and line
radiative transfer programs as in the case of spherical models but
discretizing the cloud onto a 128$^3$ cartesian grid. When
photoelectric heating is not considered, the absorption of line
radiation from the other core increases the temperatures between the
cores. The effect on $T_{\rm gas}$ is $\sim 1$ degree at the surface
between the two cores. When UV heating is included, the mutual
shielding becomes important and $T_{\rm gas}$ is reduced by up to 3\,K
between the cores. Compared to the gravitational attraction between
the spheres, the force exerted by the resulting pressure asymmetry is
of the order of one percent (assuming the change of $\Delta T \sim
3$\,K affects a few percent of the surface area). However, if the
spheres were to partially coalesce, the affected area and the
temperature asymmetry would both increase making the effect
potentially even dynamically important.

\section{Discussion} \label{sect:conclusions}

We have modelled the gas and dust temperature of dense clouds. The
results emphasize the difference between the $T_{\rm gas}$
distributions obtained with consistent radiative transfer calculation
and those resulting from the blind application of LVG model results.
If LVG calculations are used, also the increased photon escape
probability near cloud surface must be taken into account. The effect
can be several degrees and this will have consequences for chemical
models and the interpretation of observations. The photoelectric
heating is capable of raising the surface temperature significantly
but only if the cloud is shielded by less than $A_{\rm V}\sim 1^{\rm
m}$ of extinction. However, the indirect impact of this heating is
felt well beyond the region directly penetrated by UV photons. The
details of the radial temperature profile will depend on the
abundances of the outer cloud layers and the external radiation field,
both in terms of line radiation and the UV flux. Further studies
coupling the chemistry and the modelling of the thermal balance are
clearly needed.

We examined the effects on $T_{\rm gas}$ resulting from such spatial
variation of molecular abundances, velocity field, and dust grain size
distribution that are expected in dense cores. Each factor alone can
change $T_{\rm gas}$ by $\sim$1\,K or more. In the core, a strong
depletion of molecules and the reduction of turbulent motions is
capable of rising the temperature by several degrees. As pointed out
by \cite{Goldsmith01}, the effect of depletion will become less
important at higher densities when $\Lambda_{g,d}$ dominates over line
cooling. The same applies to any effect resulting from the velocity
field. However, the increase of grain sizes will significantly
decrease the coupling between gas and dust. When $T_{\rm dust}>T_{\rm
gas}$ and the density is close to $10^5$\,cm$^{-3}$, this can
compensate some of the temperature increase predicted for the inner
core. There is no observational evidence of a temperature increase at
the centre of quiescent cores but also this possibility should be
considered when interpreting observations. In more opaque clouds
(especially in conjunction with dust coagulation) $T_{\rm dust}$ will
be reduced below $T_{\rm gas}$ and, at high enough densities, will
eventually force gas temperature down at the centre of starless cores.
An increase in the grain sizes can shift this transition to densities
higher than usually assumed. However, because of the long time scale
of dust coagulation \citep[e.g.][]{Ormel09}, the effect is often
likely to be smaller than in our model.

The value of $T_{\rm gas}$ is particularly uncertain at the cloud
surface where the photoelectric heating and the dissociation of CO 
produce strong temperature gradients. However, for the dense clouds
the largest source of uncertainty still appears to be the rate of
cosmic ray heating. In the case of the one solar mass Bonnor-Ebert
sphere, a factor of five increase in $\zeta$ would raise the central
temperature by five degrees to $T_{\rm gas}\sim$12.5\,K. This may
already be excluded by direct observational evidence of much lower gas
temperatures in dense clouds \citep{Crapsi07,Harju08}. Nevertheless,
the theoretical prediction of the temperatures -- and temperature
profiles -- of dense cores still contains significant uncertainty.

\begin{figure}
\plotone{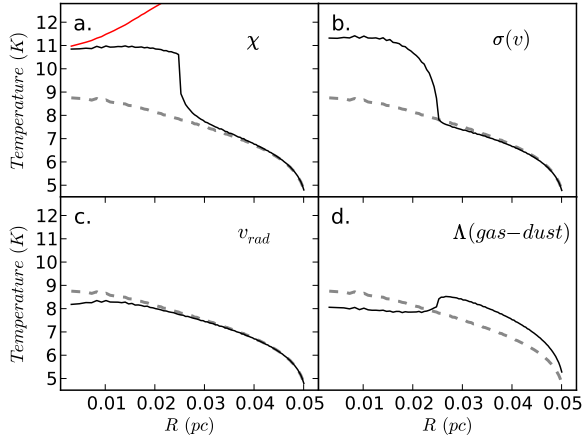}
\caption{
$T_{\rm gas}$ in a one solar mass Bonnor-Ebert sphere. The dashed line
shows the default solution with constant fractional abundances $\chi$,
a turbulent line width of $\sigma_{\rm V}=1$\,km\,s$^{-1}$, and no
large scale velocity gradients. Part of the dust temperature profile
is shown in the first frame ($T_{\rm dust}>$11\,K). Each frame shows
the effect of a single modification: a lower fractional abundance for
$r<0.025$\,pc ({\em frame a}), lower turbulence $\sigma_{\rm
V}=0.1$\,km\,s$^{-1}$ for $r<0.025$\,pc ({\em frame b}), an infall
velocity ({\em frame c}), and a lower value of $\Lambda_{g,d}$ ({\em
frame d}). See text for details.
\label{fig:BE_1.0}}
\end{figure}

\begin{figure}
\plotone{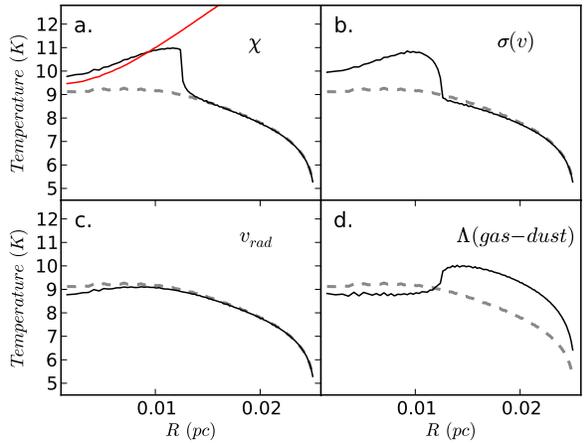}
\caption{
$T_{\rm gas}$ in 0.5\,$M_{\sun}$ Bonnor-Ebert models (lines as in
Fig.~\ref{fig:BE_1.0}).
\label{fig:BE_0.5}}
\end{figure}

\begin{figure}
\plotone{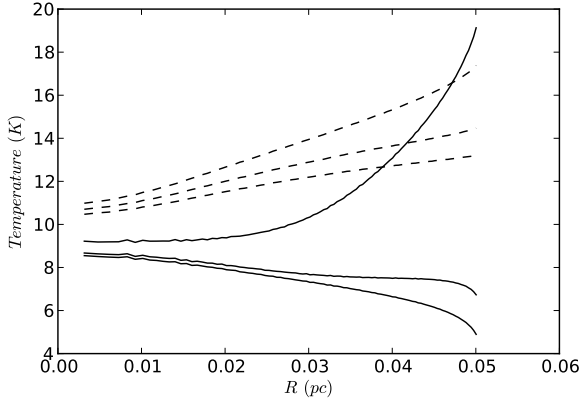}
\caption{
The radial profiles of $T_{\rm gas}$ (solid lines) and $T_{\rm dust}$ 
(dashed lines) for the one solar mass Bonnor-Ebert model, taking into
account the photoelectric heating. Reading from the top, the lines
correspond to clouds shielded by an $A_{\rm V}$ equal to 0, 1, or 2
magnitudes.
\label{fig:PEH}}
\end{figure}

\begin{figure}
\plotone{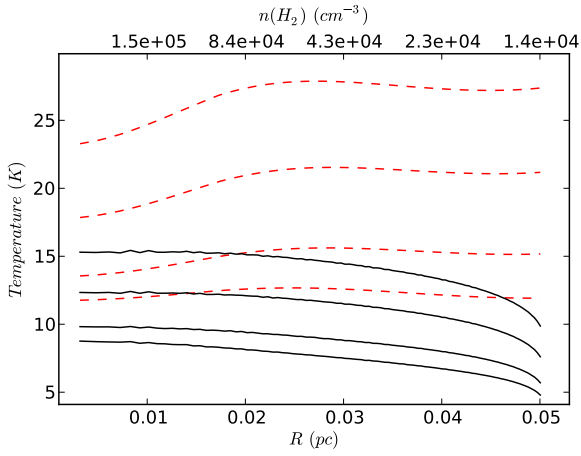}
\caption{
The radial profiles of $T_{\rm gas}$ in 1.0\,$M_{\sun}$ Bonnor-Ebert models where the
cosmic ray heating $\Gamma_{gas, cr}$ is scaled by factors 1.0, 2.0,
5.0, and 10.0 (the curves reading from the bottom). The dashed lines
are temperatures calculated using the \citet{Goldsmith01}
parameterization of $\Lambda_{gas}$ that applies to LVG models
with higher column density.
\label{fig:CR}}
\end{figure}

\begin{figure}
\plotone{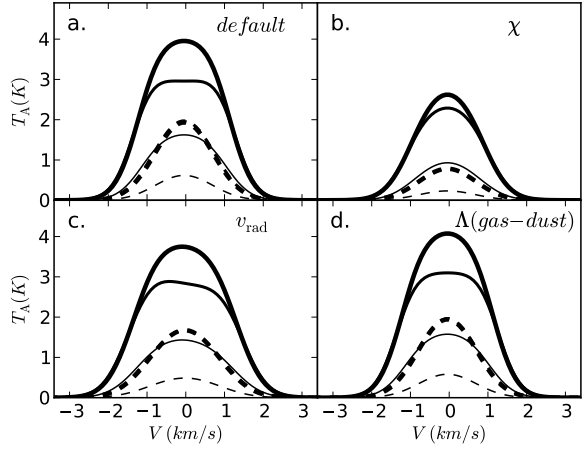}
\caption{
The $^{13}$CO spectra (solid lines) and the C$^{18}$O spectra (dashed
lines) for selected models from Fig.~\ref{fig:BE_1.0}. The curves
correspond to transitions $J=1-0$ (thick lines), $J=2-1$ (medium
line), and $J=3-2$ (thin lines). The C$^{18}$O spectra have been
scaled by a factor of two.
\label{fig:spectra_1.0}
}
\end{figure}

\begin{figure}
\plotone{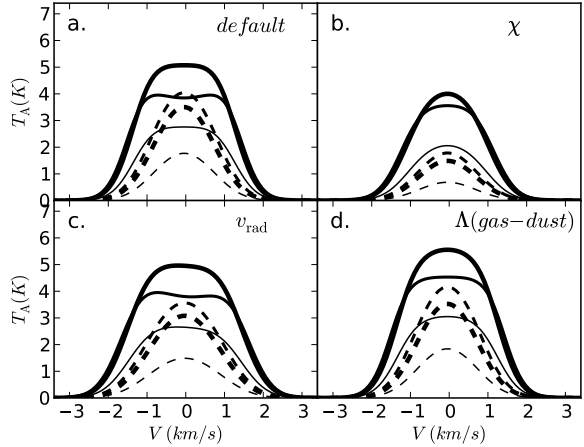}
\caption{
The $^{13}$CO and C$^{18}$O spectra for the 0.5\,$M_{\sun}$ models of
Fig.~\ref{fig:BE_0.5}. The lines are as in
Fig.~\ref{fig:spectra_1.0}.
\label{fig:spectra_0.5}
}
\end{figure}

\begin{figure}
\plotone{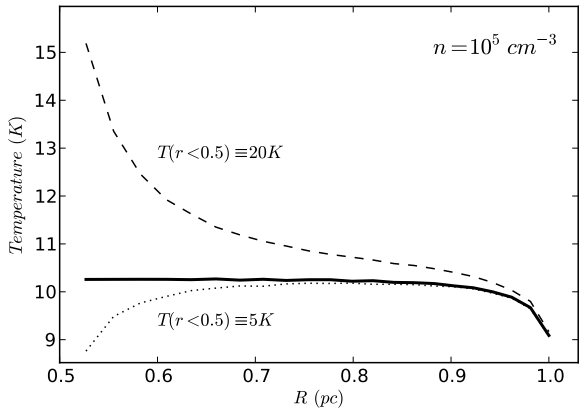}
\caption{
For the $n=10^5$\,cm$^{-3}$ model, the radial temperature profile in
the outer part of the cloud. The solid curve is the solution from
Fig.~\ref{fig:goldsmith}. The dashed line and the dotted line indicate the
temperature when the inner part of the core is forced to LTE with a
temperature of 20\,K or 5\,K, respectively.
\label{fig:LTE}}
\end{figure}

\begin{figure}
\epsscale{.99}
\plotone{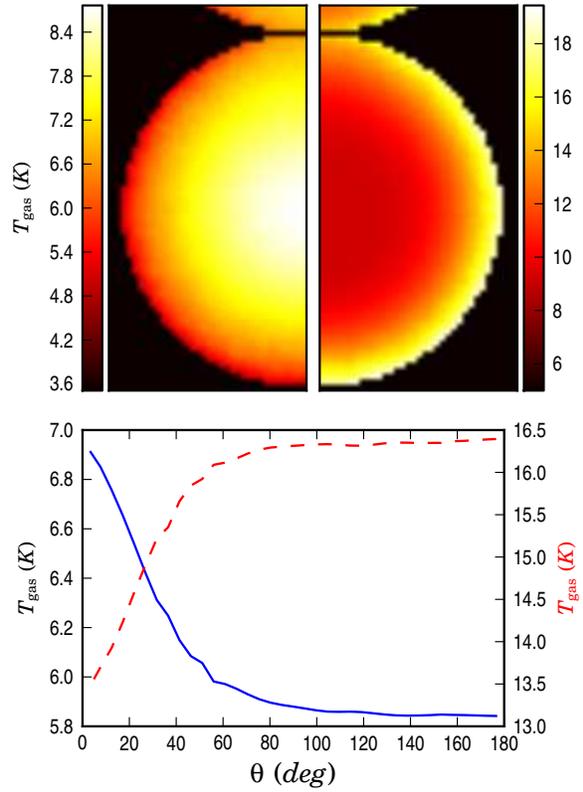}
\caption{
Gas temperature in a system consisting of two one solar mass
Bonnor-Ebert spheres that are touching each other. The images on the
top show $T_{\rm gas}$ in a cross section through the centre of the
lower sphere, the symmetry axis being vertical. The left and right
hand sides correspond, respectively, to the situation without and with
photoelectric heating. The lower frame shows the surface temperature,
averaged over 90--95\% of the core radius as the function of the angle
$\theta$. The angle is measured from the symmetry axis, as seen from
the centre of a core, and the value of $\theta=0$\degr\, corresponds
to the direction towards the other sphere. The dashed line and the
right hand axis show $T_{\rm gas}$ with photoelectric heating
included.
\label{fig:two_spheres}}
\end{figure}

The observed temperature gradients will not strongly modify the radial
density distribution of cores nor significantly affect the core
stability \citep[e.g.][]{Harju08, Galli02}. However, in the critically
stable Bonnor-Ebert models of one solar mass, the difference between
{\em isothermal} temperatures of 8\,K and 10\,K corresponds to a
factor of two increase in the central density. This shows that even
small temperature changes are important in theoretical studies.
In Figs.~\ref{fig:BE_1.0} and \ref{fig:BE_0.5} we examined separately
the effect of various parameters on $T_{\rm gas}$. In more dense
cores, with $T_{\rm dust} < T_{\rm gas}$, the effects of the $\chi$,
$\sigma_{\rm v}$, and $\Lambda_{g,d}$ parameters could accumulate,
making the temperature gradients more pronounced. If gas is not
coupled to dust, $T_{\rm gas}$ could in the central parts remain
several degrees above the temperature of the outer cloud. However, in
the centre of the 0.5\,$M_{\sun}$ model cloud the gas temperature was
already largely determined by the gas-dust coupling. In that case the
question of the dust properties becomes important because different
dust models can, in the cloud centre, lead to $T_{\rm dust}$ values
differing by more than one degree (e.g. our $T_{\rm dust}$ vs. the
\citep{Galli02} models).
The case of the two Bonnor-Ebert spheres in Sect.~\ref{sect:nonlocal}
suggests that even small temperature anisotropies may sometimes play a
role in the long term evolution of clouds. Further studies are also
required to find out how the early evolution of spherical cores is
modified relative to the isothermal case.

However, the main importance of small $T_{\rm gas}$ variations may
come via chemistry. The gas temperature directly affects chemical
reaction rates and, in particular, the depletion onto dust grains.
Therefore, the precise value of the gas temperature is relevant for
interpretation of both line and continuum data. Because the collision
rates are only proportional to $\sqrt{T_{\rm gas}}$, the direct effect
on the time scales of depletion and grain mantle accumulation is
small, less than 50\% for the kind of temperature variations observed
in our models. However, the effect on steady state abundances is more
noticeable. \citet{Aikawa05} studied chemical evolution in collapsing
clouds with initial conditions close to critical Bonnor-Ebert spheres.
If the evolution was slow enough, significant depletion was observed
in the central parts of the model clouds. In particular,
\citet{Aikawa05} included a comparison of identical models (central
density $3\times 10^6$\,cm$^{-3}$) with kinetic temperatures of 10\,K,
12\,K, and 15\,K. The ice composition was found to be very sensitive
to the temperature and this was reflected in the gas phase abundances.
A difference of two degrees could modify some abundances by a factor
of two and, in the centre where the depletion becomes significant, by
an order of magnitude or even more \citep[e.g. NH$_{3}$ and
N$_2$H$^{\rm +}$, see][Figs. 2 and 6]{Aikawa05}. It is conceivable
that in some cases the depletion will be regulated by the temperature
rise that results from the decreasing line cooling.

When observations are analyzed, there is no guarantee that different
lines (e.g., different isotopomers or different transitions of the
same molecule) would originate in identical gas volume. We already
noted that, depending on the opacity of the lines, the spatial
resolution, and the radial $T_{\rm gas}$ profile, it is possible to
observe both blueshifted and redshifted spectra towards a collapsing
cloud. In the same fashion, the kinetic temperature measured, e.g.
with NH$_3$, may not be representative for other lines. This could
lead to errors that are propagated to the derived column densities.
The evaluation of these uncertainties requires simultaneous modelling
of the thermal balance, radiative transfer and, in particular, of the
chemistry.  Such a full study is beyond the scope of the present
paper.  However, we did carry out LTE analysis of the $^{13}$CO and
C$^{18}$O lines calculated for the Bonnor-Ebert models. The
temperature gradients did not affect the column density estimates by
more than 5\% and, at least in this case, the LTE analysis would
produce accurate estimates for the total $^{13}$CO and C$^{18}$O
column density.
Thus, the temperature variations examined in this paper mainly affect
our expectations of the radial abundance profiles. These are important
considerations when line data are used to estimate the central density
or temperature of a dense core. The direct implications on the
stability or dynamic evolution of the cores are probably of secondary
importance.

\acknowledgments

MJ and NY acknowledge the financial support by the Academy of Finland Grant
127015.

\clearpage

\begin{deluxetable}{lll}
\tabletypesize{\scriptsize}
\tablecaption{The list of modified parameters affecting $T_{\rm gas}$}
\tablewidth{0pt}
\tablehead{
\colhead{Parameter} &  \colhead{Default value}  &  \colhead{Range of values$^{a}$}
}
\startdata
$\Lambda_{g,d}$                 &  Eq.~\ref{eq:gd}    & 0.3--3.0 $\times$ $\Lambda_{g,d}^0$   \\
abundances $\chi$               &  \citet{Goldsmith01} Table 1 & 0.1--1.0 $\times \chi^0$  \\
infall velocity                 &  0\,km\,s$^{-1}$    & 0.0--1.0\,km\,s$^{-1}$                \\
turbulent linewidth $\sigma_v$  &  1.0\,km\,s$^{-1}$  & 0.1--1.0\,km\,s$^{-1}$ \\
rate of cosmic rays $\zeta$     &  3$\times 10^{-17}$\,s$^{-1}$ & 1.0--10.0 $\times \zeta^0$ \\ 
\enddata
\tablenotetext{a}{The default parameter value is indicated with an upper index 0}
\label{table:parameters}
\end{deluxetable}

\clearpage

\clearpage

\begin{table}
\begin{center}
\caption{LTE column density estimates derived for the model spectra of
Figs.~\ref{fig:spectra_1.0} and \ref{fig:spectra_0.5}.}
\begin{tabular}{lcccc}
\tableline\tableline
Model version & \multicolumn{2}{c}{1.0\,$M_{\sun}$ model} & \multicolumn{2}{c}{0.5\,$M_{\sun}$ model} \\
 & $N$(C$^{18}$O)$_{\rm true}^{a}$  & 
   $N$(C$^{18}$O)$_{\rm LTE}$  &
   $N$(C$^{18}$O)$_{\rm true}^{a}$  &
   $N$(C$^{18}$O)$_{\rm LTE}$  \\
 & (cm$^{-2}$) & (cm$^{-2}$) & (cm$^{-2}$) & (cm$^{-2}$) \\
\tableline
Default                   & 1.79$\times 10^{15}$ &  1.88$\times 10^{15}$   & 3.57$\times 10^{15}$ &  3.75$\times 10^{15}$   \\
Modified $\chi$           & 6.72$\times 10^{14}$ &  6.97$\times 10^{14}$   & 1.34$\times 10^{15}$ &  1.38$\times 10^{15}$   \\
Modified $v_{\rm rad}$    & 1.79$\times 10^{15}$ &  1.88$\times 10^{15}$   & 3.57$\times 10^{15}$ &  3.74$\times 10^{15}$   \\
Modified $\Lambda_{g,d}$  & 1.79$\times 10^{15}$ &  1.89$\times 10^{15}$   & 3.57$\times 10^{15}$ &  3.74$\times 10^{15}$   \\
\tableline
\end{tabular}
\tablenotetext{a}{The true, beam averaged column density of the model}
\label{table:LTE}
\end{center}
\end{table}

\clearpage

\end{document}